\documentclass[aip,apl,amsmath,amssymb,reprint]{revtex4-1}

\usepackage{graphicx}
\usepackage{dcolumn}
\usepackage{bm}

\renewcommand{\eqref}[1]{Eq.~(\ref{#1})} 
\newcommand{\figref}[1]{Fig.~\ref{#1}} 

\newcommand{\kHz}{ \ensuremath{ \text{kHz} } }
\newcommand{\MHz}{ \ensuremath{ \text{MHz} } }

\newcommand{\keffi}{ \ensuremath{\kappa_\text{eff,i}} }
\newcommand{\keffone}{ \ensuremath{\kappa_\text{eff,1}} }
\newcommand{\kefftwo}{ \ensuremath{\kappa_\text{eff,2}} }
\newcommand{\keffloss}{ \ensuremath{\kappa_\text{eff,loss}} }
\newcommand{\kLCbare}{ \ensuremath{\kappa_\text{LC,bare}} }
\newcommand{\kLCloss}{ \ensuremath{\kappa_\text{LC,loss}} }
\newcommand{\kLCtot}{ \ensuremath{\kappa_\text{LC,tot}} }
\newcommand{\kcavi}{ \ensuremath{\kappa_\text{cav,i}} }
\newcommand{\kcavone}{ \ensuremath{\kappa_\text{cav,1}} }
\newcommand{\kcavtwo}{ \ensuremath{\kappa_\text{cav,2}} }
\newcommand{\kcavint}{ \ensuremath{\kappa_\text{cav,loss}} }
\newcommand{\kcavtot}{ \ensuremath{\kappa_\text{cav,tot}} }
\newcommand{\Deff}{ \ensuremath{\Delta_\text{eff}} }
\newcommand{\wLC}{ \ensuremath{\omega_\text{LC}} }
\newcommand{\wcav}{ \ensuremath{\omega_\text{cav}} }

\begin{document}


\title{Reconfigurable re-entrant cavity for wireless coupling to an electro-optomechanical device} 



\author{T. Menke}
\email[]{tim\_menke@g.harvard.edu}
\affiliation{Department of Physics, Harvard University, Cambridge, MA 02138, USA}

\author{P. S. Burns}
\affiliation{JILA, University of Colorado and NIST, Boulder, Colorado 80309, USA}
\affiliation{Department of Physics, University of Colorado, Boulder, Colorado 80309, USA}

\author{A. P. Higginbotham}
\affiliation{JILA, University of Colorado and NIST, Boulder, Colorado 80309, USA}
\affiliation{Department of Physics, University of Colorado, Boulder, Colorado 80309, USA}
\affiliation{National Institute of Standards and Technology (NIST), Boulder, Colorado 80305, USA}

\author{N. S. Kampel}
\affiliation{JILA, University of Colorado and NIST, Boulder, Colorado 80309, USA}
\affiliation{Department of Physics, University of Colorado, Boulder, Colorado 80309, USA}

\author{R. W. Peterson}
\affiliation{JILA, University of Colorado and NIST, Boulder, Colorado 80309, USA}
\affiliation{Department of Physics, University of Colorado, Boulder, Colorado 80309, USA}

\author{K. Cicak}
\affiliation{National Institute of Standards and Technology (NIST), Boulder, Colorado 80305, USA}

\author{R. W. Simmonds}
\affiliation{National Institute of Standards and Technology (NIST), Boulder, Colorado 80305, USA}

\author{C. A. Regal}
\affiliation{JILA, University of Colorado and NIST, Boulder, Colorado 80309, USA}
\affiliation{Department of Physics, University of Colorado, Boulder, Colorado 80309, USA}

\author{K. W. Lehnert}
\affiliation{JILA, University of Colorado and NIST, Boulder, Colorado 80309, USA}
\affiliation{Department of Physics, University of Colorado, Boulder, Colorado 80309, USA}
\affiliation{National Institute of Standards and Technology (NIST), Boulder, Colorado 80305, USA}


\date{\today}

\begin{abstract}

An electro-optomechanical device capable of microwave-to-optics conversion has recently been demonstrated, with the vision of enabling optical networks of superconducting qubits.
Here we present an improved converter design that uses a three-dimensional (3D) microwave cavity for coupling between the microwave transmission line and an integrated LC resonator on the converter chip.
The new design simplifies the optical assembly and decouples it from the microwave part of the setup.
Experimental demonstrations show that the modular device assembly allows us to flexibly tune the microwave coupling to the converter chip while maintaining small loss.
We also find that electromechanical experiments are not impacted by the additional microwave cavity.
Our design is compatible with a high-finesse optical cavity and will improve optical performance.

\end{abstract}


\maketitle 




Quantum information networks will likely combine several disparate physical systems, exploiting their different advantages \cite{Wallquist2009}.
Using superconducting circuits to process and store information and optical fields to transmit it over long distances provides a promising possible realization \cite{Barends2014,Leghtas2015, Reagor2015, Ma2012, Tsang2010, Safavi-Naeini2011, Regal2010}.
A key enabling technology for such a vision is a quantum state preserving electro-optic converter \cite{Tsang2010,Safavi-Naeini2011}.
A particular challenge lies in coupling the converter to propagating modes, which can transport information between nodes in the network.
When addressing this challange, both the optical and microwave ports of the device must be considered.

Reliable optical connectivity requires easy assembly of the components and high optical stability, especially when cooling to cryogenic temperatures.
For superconducting microwave circuits, one problem lies in coupling an electrical circuit to a microwave transmission line without also introducing undesirable coupling to other modes or otherwise spoiling the system's coherence \cite{Axline2016}.
A solution is to place a chip-based, integrated circuit inside a 3D microwave cavity, which is then coupled to a transmission line.
This approach has already been used successfully for several potential components of quantum information networks, such as superconducting qubits \cite{Paik2011a,Rigetti2012}, Josephson parametric amplifiers \cite{Narla2014}, and quantum electromechanics \cite{Yuan2015,Noguchi2016}.



Microwave-optics conversion has recently been demonstrated with an electro-optomechanical device, enabling a conversion efficiency of $\sim$10\,\% for classical signals with the prospect of quantum state transfer in the future \cite{Andrews2014}.
The efficiency was limited by the optical performance, which was lower than in the best pure optomechanics experiments \cite{Peterson2016}.
A likely source of this problem was the complex device assembly and direct connection of the transducer chip to a microwave transmission line.

In this work, we overcome these difficulties by creating a more compact, more robust system in which the optical, electrical inductor-capacitor (LC) and mechanical resonators are located inside a re-entrant microwave cavity.
The modular structure of the device decouples the microwave and optical parts of the assembly.
We expect this to simplify the optical alignment and ensure higher optical stability during thermal cycling.
In addition, the device allows for a reduction of the mirror spacing from 5\,mm to 1.2\,mm, which enables a 4-fold increase of the optomechanical coupling rate \cite{Aspelmeyer2014}.

On the microwave side, the 3D cavity allows for precise control of the coupling between the transmission line and the chip.
Cryogenic measurements at 4\,K show that the coupling rate between the microwave transmission line and the superconducting LC circuit can be varied by more than one order of magnitude around the desired value, which permits high flexibility when tuning the experimental parameters.
With further cooling to millikelvin temperatures, we show that the energy decay of the LC is dominated by coupling to the transmission line - as required for quantum state preservation - with only 17\,\% of the total loss arising from dissipation. 
Finally, we show that electromechanical measurements can readily be performed with the device.
Our approach extends the advantages of 3D microwave cavities to a crucial quantum device and opens the path to integrate an electro-optic converter with other devices that use such cavities.



\begin{figure*}[t]
  \begin{minipage}[t]{.47\linewidth}
    \null
    \includegraphics[width=\textwidth]{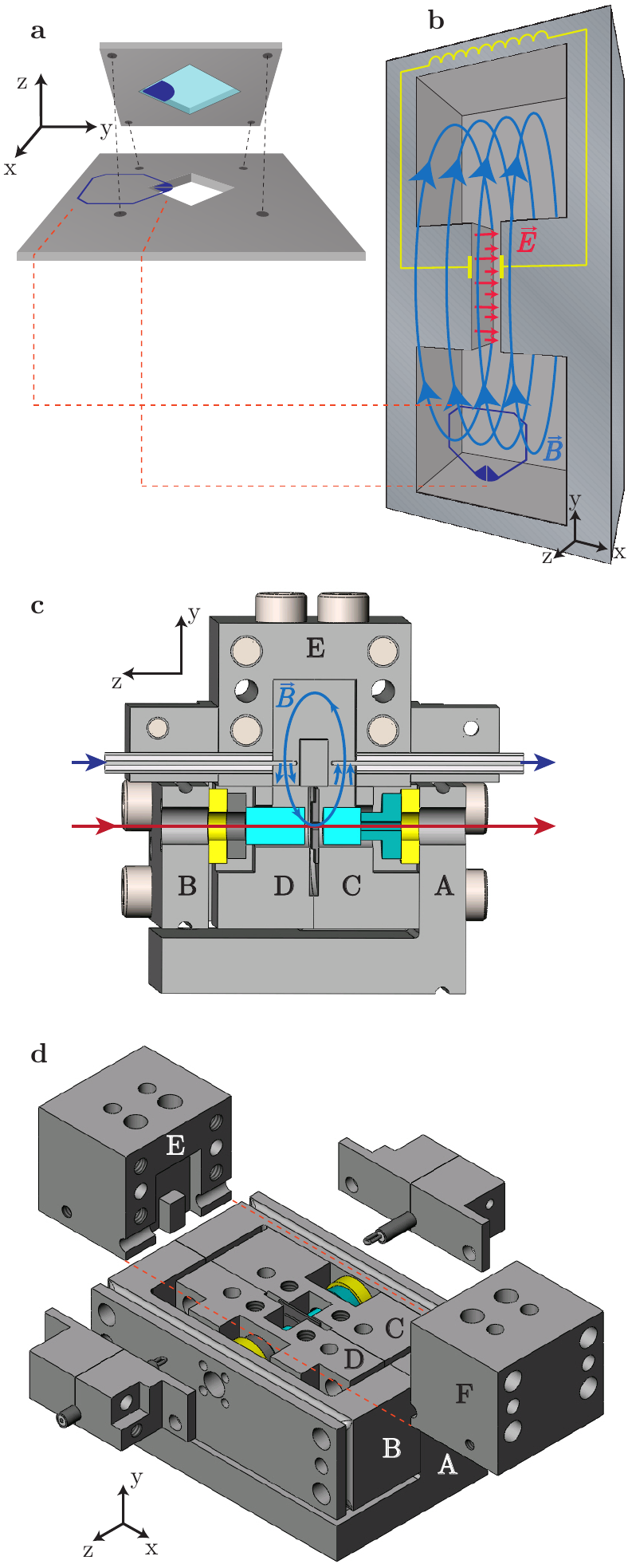}
  \end{minipage}\hfill
  \begin{minipage}[t]{0.47\textwidth}
    \caption{Overview of the converter design. (a) The microwave-to-optics transducer is assembled from two silicon chips (gray). A silicon nitride membrane (light blue) is suspended on the top chip. A niobium circuit (dark blue) is patterned on the chips. (b) Cutaway drawing of the re-entrant microwave cavity. The cavity (gray) can be modeled as an effective LC circuit (yellow line), spatially separating electric field (red) and magnetic field (blue arrows). (c) Cutaway drawing of the complete assembly. The device mount consists of several aluminum pieces, labeled A-E. The re-entrant cavity volume is formed in the middle. The cavity couples to microwave input and output (dark blue arrows) via loop antennas made from coaxial cables. The transducer chip (dark gray) is located inside an optical Fabry-P\'erot cavity. The mirrors (light blue) are brought in through holes in the microwave cavity walls. They are connected to piezoelectric crystals (yellow) via aluminum and fused-silica spacers (dark gray and turquoise, respectively). The optical path is shown as a red arrow. (d) Partially exploded view of the device, showing how the microwave and optical parts separate. The optical cavity is mounted on parts A-B. Part C serves as the sample holder for the transducer chip. At the same time, parts C-F form the microwave cavity volume. The hat consisting of parts E-F contains the re-entrant cavity's center posts and has holders for the loop antennas attached. The microwave and optical parts are in contact only via part C.}
    \label{fig:DeviceDesign}
  \end{minipage}
\end{figure*}

The microwave-to-optics transducer chip, shown in \figref{fig:DeviceDesign}a, consists of a thin membrane that couples to the electromagnetic modes of both a microwave LC resonator and an optical Fabry-P\'erot resonator.
The partially metallized portion of the membrane is part of a superconducting LC circuit and forms a mechanically compliant capacitor plate.
A displacement of the membrane shifts the resonant frequency of the circuit, which couples the mechanical and microwave modes.
In the complete conversion setup, the membrane is additionally placed inside of an optical, high-finesse Fabry-P\'erot cavity.
This forms an optomechanical system and enables bidirectional conversion between the microwave and optical fields \cite{Andrews2014}.
The work presented here, however, solely concentrates on measurements of the microwave portion of the device.

A re-entrant microwave cavity is engineered to create a reconfigurable, effective coupling between the LC circuit and the transmission line, but without coupling substantially to other modes or sources of loss.
For microwave testing purposes, the device is made out of aluminum.
For experiments that also require optical stability, we would choose copper coated invar, which minimizes thermal contraction while maintaining acceptable microwave performance.
The magnetic field of the cavity mode couples inductively to the chip circuit, as shown in \figref{fig:DeviceDesign}b.
The cavity also couples to external microwave ports via loop antennas.
The cavity can be modeled as an effective LC circuit, where the capacitance depends on the dimensions of the center posts.
By changing the post separation, we can thus control the microwave cavity frequency and effective coupling without otherwise disturbing the superconducting circuit or optical cavity.

Fig. \ref{fig:DeviceDesign}c presents the integration of the microwave cavity with an optical Fabry-P\'erot cavity.
Piezoelectric crystals are mounted outside of the microwave cavity volume.
They allow for adjustments of the optical cavity length and of the membrane position along the standing wave light field.
The exploded view in \figref{fig:DeviceDesign}d shows how the microwave and optical parts of the device have been decoupled.
This greatly reduces the complexity of the setup.
The optical cavity and membrane on the chip can now be aligned and assembled in the same way as a pure optomechanics system \cite{Purdy2012a}.
The small number of interfaces and symmetric design of the mount are expected to minimize misalignment that arises from thermal contraction.
Parts E-F, dubbed the ``hat" of the microwave cavity, contain the center posts and connect the device to the microwave circuitry via coaxial loop couplers.
We can readily change to a hat with different center post separation without interfering with any other part of the device.
The external coupling can also be adjusted by changing the size of the loop couplers.
In this way, we have precise control over the microwave cavity's resonant frequency and coupling to the external ports.



\begin{figure}
\includegraphics{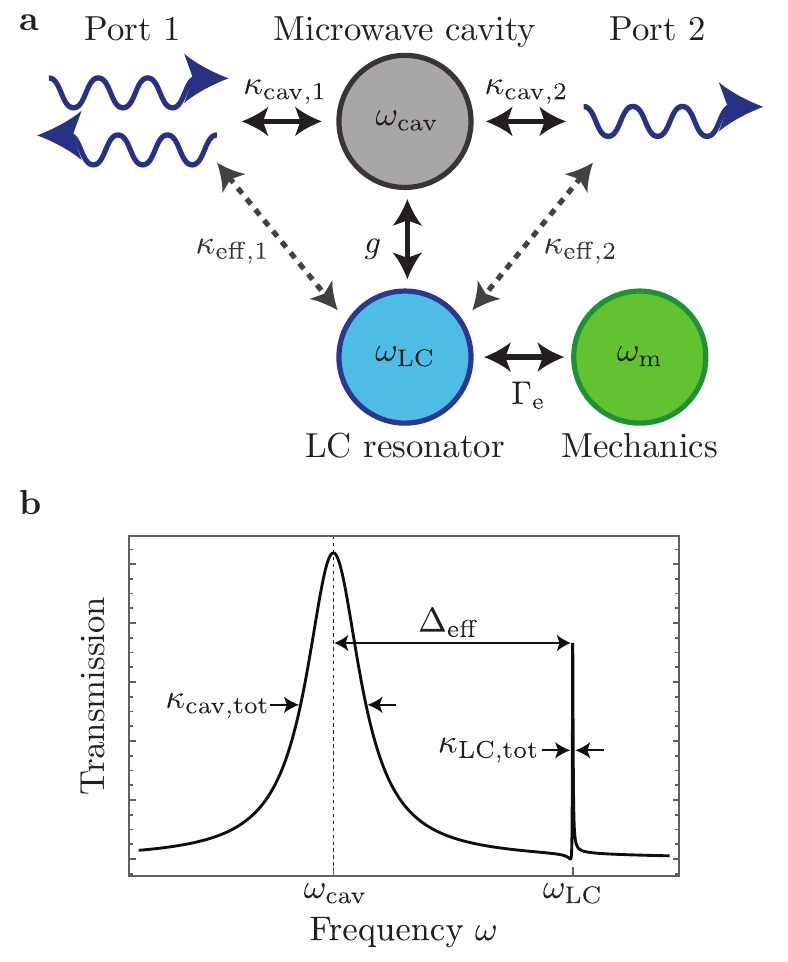}
\caption{
Theoretical description.
(a) Schematic overview of the components and coupling rates involved in the setup.
The microwave cavity (gray) couples to the transmission line (dark blue) via two ports, with external coupling rates $\kcavone$ and $\kcavtwo$, respectively.
It also couples to the LC resonator (light blue) with rate $g$.
This results in effective coupling rates $\keffone$ and $\kefftwo$ of the two ports to the LC resonator, given by \eqref{eq:keffi}.
Additionally, the microwave cavity and LC resonator have internal decay rates $\kcavint$ and $\kLCbare$, respectively (not shown).
In electromechanical measurements, the LC resonator couples to the mechanical mode of the membrane with a coupling rate $\Gamma_\text{e}$.
(b) Magnitude squared of the theoretical transfer function in absence of mechanical coupling.
Cavity linewidth $\kcavtot$, effective detuning $\Deff$, LC linewidth $\kLCtot$ labeled.
}
\label{fig:Theory}
\end{figure}

\figref{fig:Theory}a outlines the relevant coupling rates of the 3D microwave cavity and electromechanical chip.
The cavity has two ports and couples to them with external coupling rates $\kcavone$ and $\kcavtwo$, respectively.
The coupling rates are chosen such that $\kcavone \gg \kcavtwo$.
In this way, transmission and reflection measurements can be performed while minimizing the signal that is lost in reflection.
In addition, there is an internal cavity loss rate $\kcavint$ due to absorption and radiation through seams.
In total, energy stored in the cavity decays at the rate $\kcavtot = \kcavone+\kcavtwo+\kcavint$.
The inductive coupling rate between the cavity and LC is quantified by the frequency $g$.
It depends on the size and position of the LC circuit's loop inside the re-entrant cavity.
The external and inductive coupling rates can be estimated by finite element simulations to guide the design process \footnote{The electromagnetic structure solver Ansys HFSS 15.0 is used}.

The coupling of the re-entrant microwave cavity mode to both the transmission line and the LC circuit gives rise to an effective, wireless coupling of the external microwave ports to the LC circuit.
The cavity resonant frequency $\wcav$ and the LC resonator frequency $\wLC$ have been shifted from their bare values by the coupling, and their effective detuning is $\Deff = \wcav - \wLC$.
For sufficient detuning $\Deff \gtrsim \kcavtot, \, g$, the effective coupling of the LC circuit to the cavity ports is given by
\begin{equation}
    \keffi = \kcavi \frac{g^2}{ \Deff^2 + \left(\kcavtot/2\right)^2 },
    \label{eq:keffi}
\end{equation}
where $i\in\{1,2,\text{loss}\}$.
We call the couplings $\keffone$, $\kefftwo$ to the transmission line ``wireless" because there is no direct connection of the LC circuit to a waveguide or coaxial cable.
Note that the effective coupling leads to an LC dissipation rate $\kLCloss = \kLCbare + \keffloss$ that is higher than the bare loss rate $\kLCbare$ of the circuit.

The design targets an effective coupling $\keffone / 2\pi$ in the 1.5-2.0\,MHz range.
In this way, the LC circuit is well overcoupled with respect to its expected internal loss rate, which in earlier cryogenic measurements at 4\,K on a similar LC circuit was measured at 0.37\,MHz \cite{Andrews2014}.
The total LC decay rate $\kLCtot = \keffone + \kefftwo + \kLCloss$ then also fulfills the resolved sideband criterion: $\kLCtot \ll 4 \omega_\text{m}$, where the chip is designed to have a mechanical resonance frequency $\omega_\text{m} / 2\pi \sim 1.5\,\MHz$.
The last inequality ensures that electro-optic conversion occurs without photon number gain, as required for quantum state preservation. 
From \eqref{eq:keffi} we understand that we need to couple the re-entrant cavity strongly to both the LC circuit and the transmission line.
In particular, the re-entrant cavity should couple to port 1 much more strongly than it couples to its internal loss such that most of the LC decay is due to its coupling to the transmission line: $\kcavone \gg \kcavint$.
Finite element simulations show that an inductive coupling $g^\text{theo} /2\pi = 60$\,MHz and external coupling $\kcavone^\text{theo} /2\pi = 150\,\MHz$ to port 1 can be achieved.
We know from \eqref{eq:keffi} that the desired effective coupling then requires a detuning $\Deff^\text{theo} /2\pi \sim 600\,\MHz$.

\figref{fig:Theory}b shows the theoretical transmission spectrum of the device.
The cavity and LC resonator appear as peaks at their respective resonant frequencies.
The full width at half maximum (FWHM) of the Lorentzian peak centered around the cavity frequency is given by the total cavity loss rate $\kcavtot$.
The second peak at frequency $\wLC$ results from the wireless coupling.
It is much narrower and has an asymmetry that depends on the cavity-LC coupling rate $g$.
The FWHM is approximately given by the total LC decay rate $\kLCtot$.



In order to test the wireless microwave coupling concept and its compatibility with electro-optomechanics experimentally, we measure the microwave response of the device.
Furthermore, we demonstrate its modular, reconfigurable nature by measuring this response with several different microwave cavity hats that have different center post separations.
In these measurements, we test the accuracy of \eqref{eq:keffi} as a model of the device's behavior and determine its model parameters.
In particular, these measurements reveal the desirable coupling and undesirable loss that the LC inherits from the cavity.
Finally, we show that the re-entrant cavity does not prevent or otherwise interfere with the electromechanical phenomena by observing parametric interaction between the mechanical resonator's motion and the energy stored in the LC.

\begin{figure}
\includegraphics{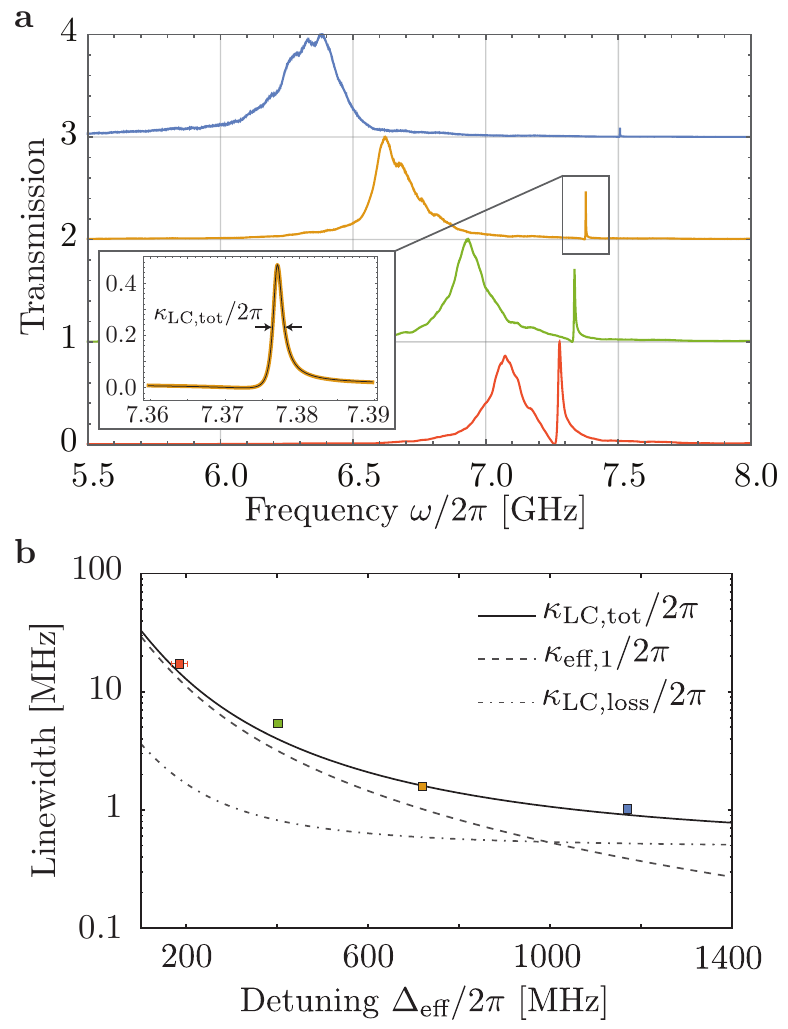}
\caption{
    (a) Normalized transmitted power versus frequency for four different center post separations: 238\,$\mu$m (blue), 270\,$\mu$m (orange), 300\,$\mu$m (green), 316\,$\mu$m (red). The broad peak corresponds to the cavity resonance, and the narrow peak corresponds to the LC resonance.
    Inset: Zoom of LC lineshape for 270\,$\mu$m post separation, with FWHM, $\kLCtot$, labeled. Fit to theoretical transfer function shown in black. Averaging over fits to all four curves gives the cavity-LC coupling $g/2\pi = 57(2)\,\MHz$, and the bare LC loss rate $\kLCbare/2\pi = 0.48(3)\,\MHz$.
    (b) LC linewidth versus effective detuning. Colored squares indicate the $\kLCtot$ obtained for the different post separations. Lines show theoretical predictions for the total LC linewidth ($\kLCtot$, solid), effective coupling ($\keffone$, dashed) and loss rate ($\kLCloss$, dot-dashed) with $g$ and $\kLCbare$ fixed from (a). The measured range of effective couplings contains the target regime 1.5-2.0\,MHz.
}
\label{fig:QuiverResults}
\end{figure}

We determine the model parameters $g$, $\kLCbare$ by measuring the device in a cryostat at 4\,K, where the niobium thin film of the LC resonator is superconducting.
Microwave cavity transmission is measured for four different hats and the resulting power spectra are plotted in \figref{fig:QuiverResults}a.
By fitting these curves to the theoretical transfer function, we extract the parameter averages $g/2\pi = 57(2)\,\MHz$ and $\kLCbare/2\pi = 0.48(3)\,\MHz$.
Note that $g/2\pi$ agrees well with the simulated coupling of 60\,MHz between cavity and LC.

We next establish that \eqref{eq:keffi} accurately models the device's behavior.
\figref{fig:QuiverResults}b shows the LC linewidths $\kLCtot$ that are extracted directly from the measured transmission spectra.
These are compared to a plot of the theoretical total loss rate $\kLCtot$ that is obtained from \eqref{eq:keffi} using the fitted model parameters.
We observe that both agree well in the regime of large detuning.
This allows us to infer the desirable and undesirable LC loss, which are plotted as $\keffone$ and $\kLCloss$ in the figure.
Note that the wireless coupling has been varied by over an order of magnitude by using the different hats.
In the target regime of 1.5-2.0 MHz for the effective coupling, the unwanted dissipation rate $\kLCloss$ constitutes at most 30\,\% of the total LC loss rate.
In a separate measurement, we find that the dissipation contribution is reduced to 17\,\% when operating the device in the millikelvin regime.

\begin{figure}
\includegraphics{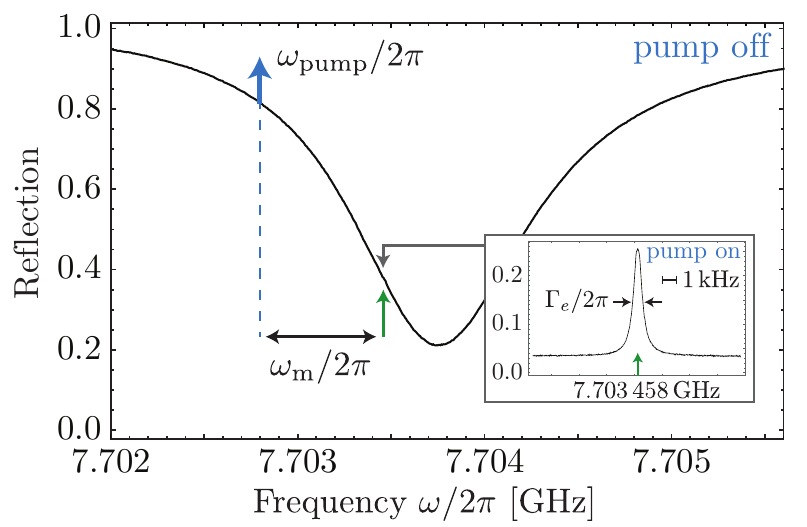}
\caption{
    Demonstration of electromechanical coupling.
    The LC resonator appears as a dip in the normalized reflected power.
    A coherent pump tone is applied below the LC resonance, at frequency $\omega_\text{pump}/2\pi$ (blue arrow).
    As a result, a peak is observed in the spectrum at the frequency of the green arrow (shown in inset).
    The frequency $\omega_\text{m}$ of the mechanical resonator is given by the frequency difference between pump and mechanics peak.
    The FWHM of the peak corresponds to the electromechanical coupling strength, which we find to be $\Gamma_\text{e} = 0.9\,\kHz$ for the applied pump power. Note that the presence of the pump shifts and deepens the LC resonance.
}
\label{fig:Mechanics}
\end{figure}

Finally, we demonstrate electromechanical interaction between the mechanical and LC resonators in the millikelvin environment of a dilution refrigerator.
A coherent pump tone is applied at a frequency that is detuned from the LC resonator by approximately the mechanical frequency.
The microwave cavity is then probed in reflection off the strongly coupled port.
\figref{fig:Mechanics} shows the reflection spectrum with the pump tone turned off.
The LC resonator appears as a dip, in contrast to a peak in transmission.
In the presence of the pump tone, the interaction between the mechanical resonator's motion and the photons stored in the LC circuit gives rise to a narrow peak on top of this feature.
This phenomenon is commonly referred to as optomechanically induced transparency \cite{Weis2010, Teufel2011a}.
For the LC circuit presented here, the pump tone leads to unexpected additional modes in the reflection spectrum.
They likely are a result of weak glue joints between the silicon chips forming the microwave-to-optics transducer.
In the \figref{fig:Mechanics} inset, we show the peak of an example mechanical mode at a frequency of 0.66\,MHz.
The electromechanical coupling rate of this mode to the LC resonator is lower than that of the desired 1.5\,MHz mode.
Therefore, a higher pump power has to be applied in order to observe the electromechanical interaction.
This leads to a deeper LC resonance that is shifted down in frequency.
With the pump turned on, the mechanics peak is centered on the shifted LC resonant frequency and displays a symmetric line shape.
We have subsequently tested other chips that exhibit only the expected mode spectrum and the higher electromechanical coupling rate of the desired 1.5\,MHz mode.
For the transducer chip and pump tone chosen for this experiment, we demonstrate an electromechanical coupling strength of $\Gamma_\text{e}/2\pi = 0.9\,\kHz$.
This shows that the basic functionality of the electromechanical system is not altered by the additional microwave cavity.



In conclusion, we have overcome an important challenge of a microwave-to-optics converter design by placing the components inside a re-entrant microwave cavity.
We were able to simplify the optical cavity assembly while achieving the desired performance of the microwave LC resonator.
Going forward, we will test the full device including a functional optomechanical part.
This device will be operated at mK temperatures in a dilution refrigerator, where we expect to enable microwave-to-optics conversion in the quantum regime.
This constitutes an important step towards a quantum microwave-optics transducer that can readibly be integrated with other components of quantum information networks.


This work was supported by the AFOSR MURI under grant number FA9550-15-1-0015, and the National Science Foundation under grant number 1125844.


\bibliography{library.bib}

\begin{thebibliography}{21}%
\makeatletter
\providecommand \@ifxundefined [1]{%
 \@ifx{#1\undefined}
}%
\providecommand \@ifnum [1]{%
 \ifnum #1\expandafter \@firstoftwo
 \else \expandafter \@secondoftwo
 \fi
}%
\providecommand \@ifx [1]{%
 \ifx #1\expandafter \@firstoftwo
 \else \expandafter \@secondoftwo
 \fi
}%
\providecommand \natexlab [1]{#1}%
\providecommand \enquote  [1]{``#1''}%
\providecommand \bibnamefont  [1]{#1}%
\providecommand \bibfnamefont [1]{#1}%
\providecommand \citenamefont [1]{#1}%
\providecommand \href@noop [0]{\@secondoftwo}%
\providecommand \href [0]{\begingroup \@sanitize@url \@href}%
\providecommand \@href[1]{\@@startlink{#1}\@@href}%
\providecommand \@@href[1]{\endgroup#1\@@endlink}%
\providecommand \@sanitize@url [0]{\catcode `\\12\catcode `\$12\catcode
  `\&12\catcode `\#12\catcode `\^12\catcode `\_12\catcode `\%12\relax}%
\providecommand \@@startlink[1]{}%
\providecommand \@@endlink[0]{}%
\providecommand \url  [0]{\begingroup\@sanitize@url \@url }%
\providecommand \@url [1]{\endgroup\@href {#1}{\urlprefix }}%
\providecommand \urlprefix  [0]{URL }%
\providecommand \Eprint [0]{\href }%
\providecommand \doibase [0]{http://dx.doi.org/}%
\providecommand \selectlanguage [0]{\@gobble}%
\providecommand \bibinfo  [0]{\@secondoftwo}%
\providecommand \bibfield  [0]{\@secondoftwo}%
\providecommand \translation [1]{[#1]}%
\providecommand \BibitemOpen [0]{}%
\providecommand \bibitemStop [0]{}%
\providecommand \bibitemNoStop [0]{.\EOS\space}%
\providecommand \EOS [0]{\spacefactor3000\relax}%
\providecommand \BibitemShut  [1]{\csname bibitem#1\endcsname}%
\let\auto@bib@innerbib\@empty
\bibitem [{\citenamefont {Wallquist}\ \emph {et~al.}(2009)\citenamefont
  {Wallquist}, \citenamefont {Hammerer}, \citenamefont {Rabl}, \citenamefont
  {Lukin},\ and\ \citenamefont {Zoller}}]{Wallquist2009}%
  \BibitemOpen
  \bibfield  {author} {\bibinfo {author} {\bibfnamefont {M.}~\bibnamefont
  {Wallquist}}, \bibinfo {author} {\bibfnamefont {K.}~\bibnamefont {Hammerer}},
  \bibinfo {author} {\bibfnamefont {P.}~\bibnamefont {Rabl}}, \bibinfo {author}
  {\bibfnamefont {M.}~\bibnamefont {Lukin}}, \ and\ \bibinfo {author}
  {\bibfnamefont {P.}~\bibnamefont {Zoller}},\ }\href@noop {} {\bibfield
  {journal} {\bibinfo  {journal} {Phys. Scr.}\ }\textbf {\bibinfo {volume}
  {T137}} (\bibinfo {year} {2009})}\BibitemShut {NoStop}%
\bibitem [{\citenamefont {Barends}\ \emph {et~al.}(2014)\citenamefont
  {Barends}, \citenamefont {Kelly}, \citenamefont {Megrant}, \citenamefont
  {Veitia}, \citenamefont {Sank}, \citenamefont {Jeffrey}, \citenamefont
  {White}, \citenamefont {Mutus}, \citenamefont {Fowler}, \citenamefont
  {Campbell}, \citenamefont {Chen}, \citenamefont {Chen}, \citenamefont
  {Chiaro}, \citenamefont {Dunsworth}, \citenamefont {Neill}, \citenamefont
  {O'Malley}, \citenamefont {Roushan}, \citenamefont {Vainsencher},
  \citenamefont {Wenner}, \citenamefont {Korotkov}, \citenamefont {Cleland},\
  and\ \citenamefont {Martinis}}]{Barends2014}%
  \BibitemOpen
  \bibfield  {author} {\bibinfo {author} {\bibfnamefont {R.}~\bibnamefont
  {Barends}}, \bibinfo {author} {\bibfnamefont {J.}~\bibnamefont {Kelly}},
  \bibinfo {author} {\bibfnamefont {A.}~\bibnamefont {Megrant}}, \bibinfo
  {author} {\bibfnamefont {A.}~\bibnamefont {Veitia}}, \bibinfo {author}
  {\bibfnamefont {D.}~\bibnamefont {Sank}}, \bibinfo {author} {\bibfnamefont
  {E.}~\bibnamefont {Jeffrey}}, \bibinfo {author} {\bibfnamefont {T.~C.}\
  \bibnamefont {White}}, \bibinfo {author} {\bibfnamefont {J.}~\bibnamefont
  {Mutus}}, \bibinfo {author} {\bibfnamefont {A.~G.}\ \bibnamefont {Fowler}},
  \bibinfo {author} {\bibfnamefont {B.}~\bibnamefont {Campbell}}, \bibinfo
  {author} {\bibfnamefont {Y.}~\bibnamefont {Chen}}, \bibinfo {author}
  {\bibfnamefont {Z.}~\bibnamefont {Chen}}, \bibinfo {author} {\bibfnamefont
  {B.}~\bibnamefont {Chiaro}}, \bibinfo {author} {\bibfnamefont
  {A.}~\bibnamefont {Dunsworth}}, \bibinfo {author} {\bibfnamefont
  {C.}~\bibnamefont {Neill}}, \bibinfo {author} {\bibfnamefont
  {P.}~\bibnamefont {O'Malley}}, \bibinfo {author} {\bibfnamefont
  {P.}~\bibnamefont {Roushan}}, \bibinfo {author} {\bibfnamefont
  {A.}~\bibnamefont {Vainsencher}}, \bibinfo {author} {\bibfnamefont
  {J.}~\bibnamefont {Wenner}}, \bibinfo {author} {\bibfnamefont {A.~N.}\
  \bibnamefont {Korotkov}}, \bibinfo {author} {\bibfnamefont {A.~N.}\
  \bibnamefont {Cleland}}, \ and\ \bibinfo {author} {\bibfnamefont {J.~M.}\
  \bibnamefont {Martinis}},\ }\href {\doibase 10.1038/nature13171} {\bibfield
  {journal} {\bibinfo  {journal} {Nature}\ }\textbf {\bibinfo {volume} {508}},\
  \bibinfo {pages} {500} (\bibinfo {year} {2014})}\BibitemShut {NoStop}%
\bibitem [{\citenamefont {Leghtas}\ \emph {et~al.}(2015)\citenamefont
  {Leghtas}, \citenamefont {Touzard}, \citenamefont {Pop}, \citenamefont {Kou},
  \citenamefont {Vlastakis}, \citenamefont {Petrenko}, \citenamefont {Sliwa},
  \citenamefont {Narla}, \citenamefont {Shankar}, \citenamefont {Hatridge},
  \citenamefont {Reagor}, \citenamefont {Frunzio}, \citenamefont {Schoelkopf},
  \citenamefont {Mirrahimi},\ and\ \citenamefont {Devoret}}]{Leghtas2015}%
  \BibitemOpen
  \bibfield  {author} {\bibinfo {author} {\bibfnamefont {Z.}~\bibnamefont
  {Leghtas}}, \bibinfo {author} {\bibfnamefont {S.}~\bibnamefont {Touzard}},
  \bibinfo {author} {\bibfnamefont {I.~M.}\ \bibnamefont {Pop}}, \bibinfo
  {author} {\bibfnamefont {A.}~\bibnamefont {Kou}}, \bibinfo {author}
  {\bibfnamefont {B.}~\bibnamefont {Vlastakis}}, \bibinfo {author}
  {\bibfnamefont {A.}~\bibnamefont {Petrenko}}, \bibinfo {author}
  {\bibfnamefont {K.~M.}\ \bibnamefont {Sliwa}}, \bibinfo {author}
  {\bibfnamefont {A.}~\bibnamefont {Narla}}, \bibinfo {author} {\bibfnamefont
  {S.}~\bibnamefont {Shankar}}, \bibinfo {author} {\bibfnamefont {M.~J.}\
  \bibnamefont {Hatridge}}, \bibinfo {author} {\bibfnamefont {M.}~\bibnamefont
  {Reagor}}, \bibinfo {author} {\bibfnamefont {L.}~\bibnamefont {Frunzio}},
  \bibinfo {author} {\bibfnamefont {R.~J.}\ \bibnamefont {Schoelkopf}},
  \bibinfo {author} {\bibfnamefont {M.}~\bibnamefont {Mirrahimi}}, \ and\
  \bibinfo {author} {\bibfnamefont {M.~H.}\ \bibnamefont {Devoret}},\ }\href
  {\doibase 10.1126/science.aaa2085} {\bibfield  {journal} {\bibinfo  {journal}
  {Science}\ }\textbf {\bibinfo {volume} {347}},\ \bibinfo {pages} {853}
  (\bibinfo {year} {2015})}\BibitemShut {NoStop}%
\bibitem [{\citenamefont {Reagor}\ \emph {et~al.}(2016)\citenamefont {Reagor},
  \citenamefont {Pfaff}, \citenamefont {Axline}, \citenamefont {Heeres},
  \citenamefont {Ofek}, \citenamefont {Sliwa}, \citenamefont {Holland},
  \citenamefont {Wang}, \citenamefont {Blumoff}, \citenamefont {Chou},
  \citenamefont {Hatridge}, \citenamefont {Frunzio}, \citenamefont {Devoret},
  \citenamefont {Jiang},\ and\ \citenamefont {Schoelkopf}}]{Reagor2015}%
  \BibitemOpen
  \bibfield  {author} {\bibinfo {author} {\bibfnamefont {M.}~\bibnamefont
  {Reagor}}, \bibinfo {author} {\bibfnamefont {W.}~\bibnamefont {Pfaff}},
  \bibinfo {author} {\bibfnamefont {C.}~\bibnamefont {Axline}}, \bibinfo
  {author} {\bibfnamefont {R.~W.}\ \bibnamefont {Heeres}}, \bibinfo {author}
  {\bibfnamefont {N.}~\bibnamefont {Ofek}}, \bibinfo {author} {\bibfnamefont
  {K.}~\bibnamefont {Sliwa}}, \bibinfo {author} {\bibfnamefont
  {E.}~\bibnamefont {Holland}}, \bibinfo {author} {\bibfnamefont
  {C.}~\bibnamefont {Wang}}, \bibinfo {author} {\bibfnamefont {J.}~\bibnamefont
  {Blumoff}}, \bibinfo {author} {\bibfnamefont {K.}~\bibnamefont {Chou}},
  \bibinfo {author} {\bibfnamefont {M.~J.}\ \bibnamefont {Hatridge}}, \bibinfo
  {author} {\bibfnamefont {L.}~\bibnamefont {Frunzio}}, \bibinfo {author}
  {\bibfnamefont {M.~H.}\ \bibnamefont {Devoret}}, \bibinfo {author}
  {\bibfnamefont {L.}~\bibnamefont {Jiang}}, \ and\ \bibinfo {author}
  {\bibfnamefont {R.~J.}\ \bibnamefont {Schoelkopf}},\ }\href
  {http://arxiv.org/abs/1508.05882} {\bibfield  {journal} {\bibinfo  {journal}
  {Phys. Rev. B}\ }\textbf {\bibinfo {volume} {94}},\ \bibinfo {pages} {1}
  (\bibinfo {year} {2016})}\BibitemShut {NoStop}%
\bibitem [{\citenamefont {Ma}\ \emph {et~al.}(2012)\citenamefont {Ma},
  \citenamefont {Herbst}, \citenamefont {Scheidl}, \citenamefont {Wang},
  \citenamefont {Kropatschek}, \citenamefont {Naylor}, \citenamefont
  {Wittmann}, \citenamefont {Mech}, \citenamefont {Kofler}, \citenamefont
  {Anisimova}, \citenamefont {Makarov}, \citenamefont {Jennewein},
  \citenamefont {Ursin},\ and\ \citenamefont {Zeilinger}}]{Ma2012}%
  \BibitemOpen
  \bibfield  {author} {\bibinfo {author} {\bibfnamefont {X.-S.}\ \bibnamefont
  {Ma}}, \bibinfo {author} {\bibfnamefont {T.}~\bibnamefont {Herbst}}, \bibinfo
  {author} {\bibfnamefont {T.}~\bibnamefont {Scheidl}}, \bibinfo {author}
  {\bibfnamefont {D.}~\bibnamefont {Wang}}, \bibinfo {author} {\bibfnamefont
  {S.}~\bibnamefont {Kropatschek}}, \bibinfo {author} {\bibfnamefont
  {W.}~\bibnamefont {Naylor}}, \bibinfo {author} {\bibfnamefont
  {B.}~\bibnamefont {Wittmann}}, \bibinfo {author} {\bibfnamefont
  {A.}~\bibnamefont {Mech}}, \bibinfo {author} {\bibfnamefont {J.}~\bibnamefont
  {Kofler}}, \bibinfo {author} {\bibfnamefont {E.}~\bibnamefont {Anisimova}},
  \bibinfo {author} {\bibfnamefont {V.}~\bibnamefont {Makarov}}, \bibinfo
  {author} {\bibfnamefont {T.}~\bibnamefont {Jennewein}}, \bibinfo {author}
  {\bibfnamefont {R.}~\bibnamefont {Ursin}}, \ and\ \bibinfo {author}
  {\bibfnamefont {A.}~\bibnamefont {Zeilinger}},\ }\href {\doibase
  10.1038/nature11472} {\bibfield  {journal} {\bibinfo  {journal} {Nature}\
  }\textbf {\bibinfo {volume} {489}},\ \bibinfo {pages} {269} (\bibinfo {year}
  {2012})}\BibitemShut {NoStop}%
\bibitem [{\citenamefont {Tsang}(2010)}]{Tsang2010}%
  \BibitemOpen
  \bibfield  {author} {\bibinfo {author} {\bibfnamefont {M.}~\bibnamefont
  {Tsang}},\ }\href {\doibase 10.1103/PhysRevA.81.063837} {\bibfield  {journal}
  {\bibinfo  {journal} {Phys. Rev. A}\ }\textbf {\bibinfo {volume} {81}},\
  \bibinfo {pages} {1} (\bibinfo {year} {2010})}\BibitemShut {NoStop}%
\bibitem [{\citenamefont {Safavi-Naeini}\ and\ \citenamefont
  {Painter}(2011)}]{Safavi-Naeini2011}%
  \BibitemOpen
  \bibfield  {author} {\bibinfo {author} {\bibfnamefont {A.~H.}\ \bibnamefont
  {Safavi-Naeini}}\ and\ \bibinfo {author} {\bibfnamefont {O.}~\bibnamefont
  {Painter}},\ }\href@noop {} {\bibfield  {journal} {\bibinfo  {journal} {New
  J. Phys.}\ }\textbf {\bibinfo {volume} {13}} (\bibinfo {year}
  {2011})}\BibitemShut {NoStop}%
\bibitem [{\citenamefont {Regal}\ and\ \citenamefont
  {Lehnert}(2011)}]{Regal2010}%
  \BibitemOpen
  \bibfield  {author} {\bibinfo {author} {\bibfnamefont {C.~A.}\ \bibnamefont
  {Regal}}\ and\ \bibinfo {author} {\bibfnamefont {K.~W.}\ \bibnamefont
  {Lehnert}},\ }\href {\doibase 10.1088/1742-6596/264/1/012025} {\bibfield
  {journal} {\bibinfo  {journal} {J. Phys. Conf. Ser.}\ }\textbf {\bibinfo
  {volume} {264}},\ \bibinfo {pages} {1} (\bibinfo {year} {2011})}\BibitemShut
  {NoStop}%
\bibitem [{\citenamefont {Axline}\ \emph {et~al.}(2016)\citenamefont {Axline},
  \citenamefont {Reagor}, \citenamefont {Heeres}, \citenamefont {Reinhold},
  \citenamefont {Wang}, \citenamefont {Shain}, \citenamefont {Pfaff},
  \citenamefont {Chu}, \citenamefont {Frunzio},\ and\ \citenamefont
  {Schoelkopf}}]{Axline2016}%
  \BibitemOpen
  \bibfield  {author} {\bibinfo {author} {\bibfnamefont {C.}~\bibnamefont
  {Axline}}, \bibinfo {author} {\bibfnamefont {M.}~\bibnamefont {Reagor}},
  \bibinfo {author} {\bibfnamefont {R.}~\bibnamefont {Heeres}}, \bibinfo
  {author} {\bibfnamefont {P.}~\bibnamefont {Reinhold}}, \bibinfo {author}
  {\bibfnamefont {C.}~\bibnamefont {Wang}}, \bibinfo {author} {\bibfnamefont
  {K.}~\bibnamefont {Shain}}, \bibinfo {author} {\bibfnamefont
  {W.}~\bibnamefont {Pfaff}}, \bibinfo {author} {\bibfnamefont
  {Y.}~\bibnamefont {Chu}}, \bibinfo {author} {\bibfnamefont {L.}~\bibnamefont
  {Frunzio}}, \ and\ \bibinfo {author} {\bibfnamefont {R.~J.}\ \bibnamefont
  {Schoelkopf}},\ }\href {\doibase 10.1063/1.4959241} {\bibfield  {journal}
  {\bibinfo  {journal} {Appl. Phys. Lett.}\ }\textbf {\bibinfo {volume}
  {109}},\ \bibinfo {pages} {042601} (\bibinfo {year} {2016})}\BibitemShut
  {NoStop}%
\bibitem [{\citenamefont {Paik}\ \emph {et~al.}(2011)\citenamefont {Paik},
  \citenamefont {Schuster}, \citenamefont {Bishop}, \citenamefont {Kirchmair},
  \citenamefont {Catelani}, \citenamefont {Sears}, \citenamefont {Johnson},
  \citenamefont {Reagor}, \citenamefont {Frunzio}, \citenamefont {Glazman},
  \citenamefont {Girvin}, \citenamefont {Devoret},\ and\ \citenamefont
  {Schoelkopf}}]{Paik2011a}%
  \BibitemOpen
  \bibfield  {author} {\bibinfo {author} {\bibfnamefont {H.}~\bibnamefont
  {Paik}}, \bibinfo {author} {\bibfnamefont {D.~I.}\ \bibnamefont {Schuster}},
  \bibinfo {author} {\bibfnamefont {L.~S.}\ \bibnamefont {Bishop}}, \bibinfo
  {author} {\bibfnamefont {G.}~\bibnamefont {Kirchmair}}, \bibinfo {author}
  {\bibfnamefont {G.}~\bibnamefont {Catelani}}, \bibinfo {author}
  {\bibfnamefont {a.~P.}\ \bibnamefont {Sears}}, \bibinfo {author}
  {\bibfnamefont {B.~R.}\ \bibnamefont {Johnson}}, \bibinfo {author}
  {\bibfnamefont {M.~J.}\ \bibnamefont {Reagor}}, \bibinfo {author}
  {\bibfnamefont {L.}~\bibnamefont {Frunzio}}, \bibinfo {author} {\bibfnamefont
  {L.~I.}\ \bibnamefont {Glazman}}, \bibinfo {author} {\bibfnamefont {S.~M.}\
  \bibnamefont {Girvin}}, \bibinfo {author} {\bibfnamefont {M.~H.}\
  \bibnamefont {Devoret}}, \ and\ \bibinfo {author} {\bibfnamefont {R.~J.}\
  \bibnamefont {Schoelkopf}},\ }\href {\doibase 10.1103/PhysRevLett.107.240501}
  {\bibfield  {journal} {\bibinfo  {journal} {Phys. Rev. Lett.}\ }\textbf
  {\bibinfo {volume} {107}},\ \bibinfo {pages} {1} (\bibinfo {year} {2011})},\
  \Eprint {http://arxiv.org/abs/1105.4652} {arXiv:1105.4652} \BibitemShut
  {NoStop}%
\bibitem [{\citenamefont {Rigetti}\ \emph {et~al.}(2012)\citenamefont
  {Rigetti}, \citenamefont {Gambetta}, \citenamefont {Poletto}, \citenamefont
  {Plourde}, \citenamefont {Chow}, \citenamefont {C{\'{o}}rcoles},
  \citenamefont {Smolin}, \citenamefont {Merkel}, \citenamefont {Rozen},
  \citenamefont {Keefe}, \citenamefont {Rothwell}, \citenamefont {Ketchen},\
  and\ \citenamefont {Steffen}}]{Rigetti2012}%
  \BibitemOpen
  \bibfield  {author} {\bibinfo {author} {\bibfnamefont {C.}~\bibnamefont
  {Rigetti}}, \bibinfo {author} {\bibfnamefont {J.~M.}\ \bibnamefont
  {Gambetta}}, \bibinfo {author} {\bibfnamefont {S.}~\bibnamefont {Poletto}},
  \bibinfo {author} {\bibfnamefont {B.~L.~T.}\ \bibnamefont {Plourde}},
  \bibinfo {author} {\bibfnamefont {J.~M.}\ \bibnamefont {Chow}}, \bibinfo
  {author} {\bibfnamefont {A.~D.}\ \bibnamefont {C{\'{o}}rcoles}}, \bibinfo
  {author} {\bibfnamefont {J.~A.}\ \bibnamefont {Smolin}}, \bibinfo {author}
  {\bibfnamefont {S.~T.}\ \bibnamefont {Merkel}}, \bibinfo {author}
  {\bibfnamefont {J.~R.}\ \bibnamefont {Rozen}}, \bibinfo {author}
  {\bibfnamefont {G.~A.}\ \bibnamefont {Keefe}}, \bibinfo {author}
  {\bibfnamefont {M.~B.}\ \bibnamefont {Rothwell}}, \bibinfo {author}
  {\bibfnamefont {M.~B.}\ \bibnamefont {Ketchen}}, \ and\ \bibinfo {author}
  {\bibfnamefont {M.}~\bibnamefont {Steffen}},\ }\href {\doibase
  10.1103/PhysRevB.86.100506} {\bibfield  {journal} {\bibinfo  {journal} {Phys.
  Rev. B}\ }\textbf {\bibinfo {volume} {86}},\ \bibinfo {pages} {1} (\bibinfo
  {year} {2012})}\BibitemShut {NoStop}%
\bibitem [{\citenamefont {Narla}\ \emph {et~al.}(2014)\citenamefont {Narla},
  \citenamefont {Sliwa}, \citenamefont {Hatridge}, \citenamefont {Shankar},
  \citenamefont {Frunzio}, \citenamefont {Schoelkopf},\ and\ \citenamefont
  {Devoret}}]{Narla2014}%
  \BibitemOpen
  \bibfield  {author} {\bibinfo {author} {\bibfnamefont {A.}~\bibnamefont
  {Narla}}, \bibinfo {author} {\bibfnamefont {K.~M.}\ \bibnamefont {Sliwa}},
  \bibinfo {author} {\bibfnamefont {M.}~\bibnamefont {Hatridge}}, \bibinfo
  {author} {\bibfnamefont {S.}~\bibnamefont {Shankar}}, \bibinfo {author}
  {\bibfnamefont {L.}~\bibnamefont {Frunzio}}, \bibinfo {author} {\bibfnamefont
  {R.~J.}\ \bibnamefont {Schoelkopf}}, \ and\ \bibinfo {author} {\bibfnamefont
  {M.~H.}\ \bibnamefont {Devoret}},\ }\href {\doibase 10.1063/1.4883373}
  {\bibfield  {journal} {\bibinfo  {journal} {Appl. Phys. Lett.}\ }\textbf
  {\bibinfo {volume} {104}},\ \bibinfo {pages} {232605} (\bibinfo {year}
  {2014})}\BibitemShut {NoStop}%
\bibitem [{\citenamefont {Yuan}\ \emph {et~al.}(2015)\citenamefont {Yuan},
  \citenamefont {Singh}, \citenamefont {Blanter},\ and\ \citenamefont
  {Steele}}]{Yuan2015}%
  \BibitemOpen
  \bibfield  {author} {\bibinfo {author} {\bibfnamefont {M.}~\bibnamefont
  {Yuan}}, \bibinfo {author} {\bibfnamefont {V.}~\bibnamefont {Singh}},
  \bibinfo {author} {\bibfnamefont {Y.~M.}\ \bibnamefont {Blanter}}, \ and\
  \bibinfo {author} {\bibfnamefont {G.~A.}\ \bibnamefont {Steele}},\ }\href
  {\doibase 10.1038/ncomms9491} {\bibfield  {journal} {\bibinfo  {journal}
  {Nat. Commun.}\ }\textbf {\bibinfo {volume} {6}},\ \bibinfo {pages} {8491}
  (\bibinfo {year} {2015})}\BibitemShut {NoStop}%
\bibitem [{\citenamefont {Noguchi}\ \emph {et~al.}(2016)\citenamefont
  {Noguchi}, \citenamefont {Yamazaki}, \citenamefont {Ataka}, \citenamefont
  {Fujita}, \citenamefont {Tabuchi}, \citenamefont {Ishikawa}, \citenamefont
  {Usami},\ and\ \citenamefont {Nakamura}}]{Noguchi2016}%
  \BibitemOpen
  \bibfield  {author} {\bibinfo {author} {\bibfnamefont {A.}~\bibnamefont
  {Noguchi}}, \bibinfo {author} {\bibfnamefont {R.}~\bibnamefont {Yamazaki}},
  \bibinfo {author} {\bibfnamefont {M.}~\bibnamefont {Ataka}}, \bibinfo
  {author} {\bibfnamefont {H.}~\bibnamefont {Fujita}}, \bibinfo {author}
  {\bibfnamefont {Y.}~\bibnamefont {Tabuchi}}, \bibinfo {author} {\bibfnamefont
  {T.}~\bibnamefont {Ishikawa}}, \bibinfo {author} {\bibfnamefont
  {K.}~\bibnamefont {Usami}}, \ and\ \bibinfo {author} {\bibfnamefont
  {Y.}~\bibnamefont {Nakamura}},\ }\href@noop {} {\bibfield  {journal}
  {\bibinfo  {journal} {New J. Phys.}\ }\textbf {\bibinfo {volume} {18}}
  (\bibinfo {year} {2016})}\BibitemShut {NoStop}%
\bibitem [{\citenamefont {Andrews}\ \emph {et~al.}(2014)\citenamefont
  {Andrews}, \citenamefont {Peterson}, \citenamefont {Purdy}, \citenamefont
  {Cicak}, \citenamefont {Simmonds}, \citenamefont {Regal},\ and\ \citenamefont
  {Lehnert}}]{Andrews2014}%
  \BibitemOpen
  \bibfield  {author} {\bibinfo {author} {\bibfnamefont {R.~W.}\ \bibnamefont
  {Andrews}}, \bibinfo {author} {\bibfnamefont {R.~W.}\ \bibnamefont
  {Peterson}}, \bibinfo {author} {\bibfnamefont {T.~P.}\ \bibnamefont {Purdy}},
  \bibinfo {author} {\bibfnamefont {K.}~\bibnamefont {Cicak}}, \bibinfo
  {author} {\bibfnamefont {R.~W.}\ \bibnamefont {Simmonds}}, \bibinfo {author}
  {\bibfnamefont {C.~A.}\ \bibnamefont {Regal}}, \ and\ \bibinfo {author}
  {\bibfnamefont {K.~W.}\ \bibnamefont {Lehnert}},\ }\href {\doibase
  10.1038/nphys2911} {\bibfield  {journal} {\bibinfo  {journal} {Nat. Phys.}\
  }\textbf {\bibinfo {volume} {10}},\ \bibinfo {pages} {321} (\bibinfo {year}
  {2014})}\BibitemShut {NoStop}%
\bibitem [{\citenamefont {Peterson}\ \emph {et~al.}(2016)\citenamefont
  {Peterson}, \citenamefont {Purdy}, \citenamefont {Kampel}, \citenamefont
  {Andrews}, \citenamefont {Yu}, \citenamefont {Lehnert},\ and\ \citenamefont
  {Regal}}]{Peterson2016}%
  \BibitemOpen
  \bibfield  {author} {\bibinfo {author} {\bibfnamefont {R.~W.}\ \bibnamefont
  {Peterson}}, \bibinfo {author} {\bibfnamefont {T.~P.}\ \bibnamefont {Purdy}},
  \bibinfo {author} {\bibfnamefont {N.~S.}\ \bibnamefont {Kampel}}, \bibinfo
  {author} {\bibfnamefont {R.~W.}\ \bibnamefont {Andrews}}, \bibinfo {author}
  {\bibfnamefont {P.-L.}\ \bibnamefont {Yu}}, \bibinfo {author} {\bibfnamefont
  {K.~W.}\ \bibnamefont {Lehnert}}, \ and\ \bibinfo {author} {\bibfnamefont
  {C.~A.}\ \bibnamefont {Regal}},\ }\href {\doibase
  10.1103/PhysRevLett.116.063601} {\bibfield  {journal} {\bibinfo  {journal}
  {Phys. Rev. Lett.}\ }\textbf {\bibinfo {volume} {116}},\ \bibinfo {pages}
  {063601} (\bibinfo {year} {2016})}\BibitemShut {NoStop}%
\bibitem [{\citenamefont {Aspelmeyer}, \citenamefont {Kippenberg},\ and\
  \citenamefont {Marquardt}(2014)}]{Aspelmeyer2014}%
  \BibitemOpen
  \bibfield  {author} {\bibinfo {author} {\bibfnamefont {M.}~\bibnamefont
  {Aspelmeyer}}, \bibinfo {author} {\bibfnamefont {T.~J.}\ \bibnamefont
  {Kippenberg}}, \ and\ \bibinfo {author} {\bibfnamefont {F.}~\bibnamefont
  {Marquardt}},\ }\href {\doibase 10.1103/RevModPhys.86.1391} {\bibfield
  {journal} {\bibinfo  {journal} {Rev. Mod. Phys.}\ }\textbf {\bibinfo {volume}
  {86}},\ \bibinfo {pages} {1391} (\bibinfo {year} {2014})}\BibitemShut
  {NoStop}%
\bibitem [{\citenamefont {Purdy}\ \emph {et~al.}(2012)\citenamefont {Purdy},
  \citenamefont {Peterson}, \citenamefont {Yu},\ and\ \citenamefont
  {Regal}}]{Purdy2012a}%
  \BibitemOpen
  \bibfield  {author} {\bibinfo {author} {\bibfnamefont {T.~P.}\ \bibnamefont
  {Purdy}}, \bibinfo {author} {\bibfnamefont {R.~W.}\ \bibnamefont {Peterson}},
  \bibinfo {author} {\bibfnamefont {P.~L.}\ \bibnamefont {Yu}}, \ and\ \bibinfo
  {author} {\bibfnamefont {C.~A.}\ \bibnamefont {Regal}},\ }\href@noop {}
  {\bibfield  {journal} {\bibinfo  {journal} {New J. Phys.}\ }\textbf {\bibinfo
  {volume} {14}} (\bibinfo {year} {2012})}\BibitemShut {NoStop}%
\bibitem [{Note1()}]{Note1}%
  \BibitemOpen
  \bibinfo {note} {The electromagnetic structure solver Ansys HFSS 15.0 is
  used}\BibitemShut {NoStop}%
\bibitem [{\citenamefont {Weis}\ \emph {et~al.}(2010)\citenamefont {Weis},
  \citenamefont {Rivi{\`{e}}re}, \citenamefont {Del{\'{e}}glise}, \citenamefont
  {Gavartin}, \citenamefont {Arcizet}, \citenamefont {Schliesser},\ and\
  \citenamefont {Kippenberg}}]{Weis2010}%
  \BibitemOpen
  \bibfield  {author} {\bibinfo {author} {\bibfnamefont {S.}~\bibnamefont
  {Weis}}, \bibinfo {author} {\bibfnamefont {R.}~\bibnamefont {Rivi{\`{e}}re}},
  \bibinfo {author} {\bibfnamefont {S.}~\bibnamefont {Del{\'{e}}glise}},
  \bibinfo {author} {\bibfnamefont {E.}~\bibnamefont {Gavartin}}, \bibinfo
  {author} {\bibfnamefont {O.}~\bibnamefont {Arcizet}}, \bibinfo {author}
  {\bibfnamefont {A.}~\bibnamefont {Schliesser}}, \ and\ \bibinfo {author}
  {\bibfnamefont {T.~J.}\ \bibnamefont {Kippenberg}},\ }\href {\doibase
  10.1126/science.1195596} {\bibfield  {journal} {\bibinfo  {journal}
  {Science}\ }\textbf {\bibinfo {volume} {330}},\ \bibinfo {pages} {1520}
  (\bibinfo {year} {2010})}\BibitemShut {NoStop}%
\bibitem [{\citenamefont {Teufel}\ \emph {et~al.}(2011)\citenamefont {Teufel},
  \citenamefont {Li}, \citenamefont {Allman}, \citenamefont {Cicak},
  \citenamefont {Sirois}, \citenamefont {Whittaker},\ and\ \citenamefont
  {Simmonds}}]{Teufel2011a}%
  \BibitemOpen
  \bibfield  {author} {\bibinfo {author} {\bibfnamefont {J.~D.}\ \bibnamefont
  {Teufel}}, \bibinfo {author} {\bibfnamefont {D.}~\bibnamefont {Li}}, \bibinfo
  {author} {\bibfnamefont {M.~S.}\ \bibnamefont {Allman}}, \bibinfo {author}
  {\bibfnamefont {K.}~\bibnamefont {Cicak}}, \bibinfo {author} {\bibfnamefont
  {A.~J.}\ \bibnamefont {Sirois}}, \bibinfo {author} {\bibfnamefont {J.~D.}\
  \bibnamefont {Whittaker}}, \ and\ \bibinfo {author} {\bibfnamefont {R.~W.}\
  \bibnamefont {Simmonds}},\ }\href {\doibase 10.1038/nature09898} {\bibfield
  {journal} {\bibinfo  {journal} {Nature}\ }\textbf {\bibinfo {volume} {471}},\
  \bibinfo {pages} {204} (\bibinfo {year} {2011})}\BibitemShut {NoStop}%
\end{thebibliography}%

\end{document}